\documentstyle[prd,aps,preprint,psfig]{revtex}
\begin{document}
\tighten
\title{Double inflation in supergravity and the primordial black hole
  formation}
\author{Toshiyuki Kanazawa} \address{Department of Physics, University
  of Tokyo, Tokyo, 113-0033, Japan} 
\author{M. Kawasaki} \address{Research Center for the Early Universe
  (RESCEU), University of Tokyo, Tokyo, 113-0033, Japan}
\author{T. Yanagida} \address{Department of Physics and RESCEU,
  University of Tokyo, Tokyo 113-0033, Japan}

\date{\today}

\maketitle

\begin{abstract}
  We study a double inflation model (a hybrid inflation $+$ a new
  inflation) in supergravity and discuss the formation of primordial
  black holes (PBHs) with mass $\sim 10^{-20}-10^{5}M_{\odot}$.
  We find that in a wide range of parameter space, we obtain PBHs
  which amount to $\Omega \simeq 1$, i.e., PBH dark matter. Also, we
  find a set of inflation parameters which produces PBHs evaporating
  now.  Those PBHs may be responsible for antiproton fluxes observed
  by the BESS experiment.
\end{abstract}


\section{Introduction}

In the framework of supergravity the reheating temperature of
inflation should be low enough to avoid overproduction of
gravitinos~\cite{Weinberg,Ellis}. The new inflation
model~\cite{New-inflation} generally predicts a very low reheating
temperature and hence it is the most attractive among many inflation
models~\cite{Linde-text}. However, the new inflation suffers from a
fine-tuning problem about the initial condition; i.e., for a
successful new inflation, the initial value of the inflaton should be
very close to the local maximum of the potential in a large region
whose size is much longer than the horizon of the universe.

A framework of a double inflation~\cite{dynamical-tuning} was proposed
to solve the initial value problem of the new inflation
model\footnote{Different models of double inflation were studied by
  various authors~\cite{doubleinf}.  }. It was shown that the above
serious problem is naturally solved by supergravity effects if there
exists a preinflation (e.g., hybrid inflation~\cite{hybrid-original})
with a sufficiently large Hubble parameter before the new
inflation~\cite{dynamical-tuning}.

In this double inflation model, density fluctuations produced by both
inflations are cosmologically relevant if the $e$-fold number of the
new inflation is smaller than $\sim 60$ (the total $e$-fold number
$\sim 60$ is required to solve flatness and horizon problems in the
standard big bang cosmology~\cite{Kolb-Turner}). In this case, the
preinflation should account for the density fluctuations on large
cosmological scales [including the Cosmic Background Explorer (COBE)
scales] while the new inflation produces density fluctuations on
smaller scales.  Although the amplitude of the fluctuations on large
scales should be normalized to the COBE data~\cite{COBE}, fluctuations
on small scales are free from the COBE normalization and can have
arbitrary power matched to observations.  In Ref.\cite{KKSY-LSS}, a
cosmological implication of the double inflation for the large-scale
structure formation was discussed. In this paper, we study primordial
black hole (PBH) formation in the double inflation model. In
Refs.~\cite{KSY-PBH,KY-PBH}, the production of black hole MACHO was
investigated in the double inflation model for a special
case\footnote{Different models for the PBH formation have been studied
  in Ref.~\cite{otherPBH}.}. Here, we consider a wide range of
parameter space where PBHs are formed\footnote{In this paper we
  investigate the PBHs with mass $\sim 10^{-20} - 10^{5} M_{\odot}$.
  The upper bound of $10^5 M_{\odot}$ is not rigorous but much heavier
  black holes can be produced in the present model if we take the
  appropriate model parameters. However, the mass of the black holes
  should be less than the galactic mass $\sim 10^{12} M_{\odot}$,
  otherwise the power spectrum conflicts the observations ( e.g.
  distribution of galaxies ).  The lower bound $\sim 10^{-20}
  M_{\odot}$ comes from the requirement that $e$-fold number of the
  new inflation should be larger than $0$.}. In particular, we show
that the double inflation creates small PBHs evaporating now if those
PBHs are produced during matter-dominated (MD) era, i.e., before the
end of reheating process after the new inflation. We stress that these
evaporating PBHs may account for antiproton fluxes observed by the
BESS experiment~\cite{BESS}.

Throughout this paper the gravitational scale ($\simeq 2.4 \times
10^{18}$ GeV) is taken to be unity.

\section{Black hole formation}

In a radiation-dominated (RD) universe, PBHs are formed if the density
fluctuations $\delta $ at horizon crossing satisfy a condition $1/3
\le \delta \le 1$~\cite{Carr,Green}, where $\delta$ is the over
density at the horizon scale. Masses of the black holes $M$ are
roughly equal to the horizon mass,
\begin{equation}
    \label{bh-mass}
    M \simeq 4\pi\sqrt{\frac{3}{\rho}}
    \simeq 0.066 M_{\odot} \left(\frac{T}{{\rm GeV}}\right)^{-2}
    \left( \frac{g_{*}}{50} \right)^{{-1/2}},
\end{equation}
where $\rho$, $T$, and $g_{*}$ are the total cosmic density,
temperature, and statistical degrees of freedom at the horizon
crossing, respectively.

The horizon length at the black hole formation epoch ($T=T_*$)
corresponds to the scale $L_{*}$ in the present universe given by 
\begin{equation}
    \label{bh-scale}
    L_{*} \simeq \frac{a(T_0)}{a(T_{*})}H^{-1}(T_{*})
    \simeq 0.064~ {\rm pc} \left( \frac{T_*}{\rm GeV} \right)^{{-1}}
    \left( \frac{g_{*}}{50} \right)^{{-1/6}},
\end{equation}
where $T_{0}$ is the temperature of the present universe. The comoving
wave number corresponding to this length scale, $k_{*} \equiv 2\pi/
L_{*}$, is
\begin{equation}
  k_{*} \simeq 1.0\times 10^{8} {\rm Mpc}^{-1} \left( \frac{g_{*}}{50}
  \right)^{1/6} \left( \frac{T_{*}}{\rm GeV} \right) .
\end{equation}
Thus, we can write the PBH mass as a function of comoving wave number
as
\begin{equation}
\label{mass-scale-RD}
M_{*}\simeq 6.4\times 10^{14} M_{\odot} \left( \frac{g_{*}}{50}
\right)^{-1/6} \left( \frac{k_{*}}{{\rm Mpc}^{-1}}\right)^{-2}.
\end{equation}

The mass fraction $\beta_{*} (= \rho_{BH}/\rho)$ of PBHs of mass
$M_{*}$ is given by~\cite{Green}
\begin{equation}
    \label{bh-frac}
    \beta_{*}(M_{*}) = \int_{1/3}^{1} 
    \frac{d\delta}{\sqrt{2\pi}\sigma(M_*)}
    \exp\left(-\frac{\delta^2}{2\sigma^2(M_{*})}\right)
    \simeq \sigma(M_{*}) 
    \exp\left(-\frac{1}{18\sigma^2(M_{*})}\right) ,
\end{equation}
where $\sigma(M_{*})$ is the mass variance at the horizon crossing.
Notice that the mass fraction $\beta_{*}(M_{*})$ drops off sharply as
$\sigma(M)$ decreases. The density of the black holes of mass $M_{*}$,
$\rho_{BH}(M_{*})$, is given by
\begin{equation}
    \label{bh-density}
    \frac{\rho_{BH}(M_{*})}{s} \simeq \frac{3}{4}\beta_{*}(M_{*})T_{*},
\end{equation}
where $s$ is the entropy density.  Since $\rho_{BH}/s$ is constant at
$T<T_{*}$, we estimate the density parameter $\Omega_{BH}(M_{*})$ of
the black holes in the present universe as
\begin{equation}
    \Omega_{BH}(M_{*})h^2 \simeq 2.1\times 10^8 \beta_{*}(M_{*}) \left(
\frac{T_*}{\rm GeV} \right),
\end{equation}
where $h$ is the present Hubble constant in units of 100~km/sec/Mpc.
We write it as a function of PBH mass or PBH scale as
\begin{equation}
\label{Omega-RD-mass}
   \Omega_{BH}(M_{*})h^2 \simeq 5.4\times 10^{7} \beta_{*}(M_{*}) \left
     ( \frac{g_{*}}{50} \right)^{-1/4} \left( \frac{M_*}{M_{\odot}}
     \right)^{-1/2},
\end{equation}
or
\begin{equation}
  \Omega_{BH}(M_{*})h^2 \simeq 2.1 \beta_{*}(M_{*}) \left (
    \frac{g_{*}}{50} \right)^{-1/6} \left( \frac{k_{*}}{{\rm
        Mpc}^{-1}} \right).
\end{equation}

As for the mass of the PBHs produced during the RD era, we have a
lower limit $M_{R}$. The mass of PBH produced at the reheating epoch
is given by [see Eq.(\ref{bh-mass})]
\begin{equation}
M_R \simeq 0.066 M_\odot \left( \frac{T_R}{\rm GeV} \right)^{-2} \left
  ( \frac{g_{*}}{50} \right)^{-1/2},
\end{equation}
where $T_{R}$ is the reheating temperature. As seen later $T_{R}$ is
less than $10^{6}$GeV in our inflation model, and hence $M_{R}$ is
larger than $\sim 10^{-13}M_{\odot}$. Thus, the PBHs lighter than
$M_{R}$ should be produced during the MD era.

In a MD universe, a relation between the comoving scale $L_{*}$ and
the horizon mass $M_{*}$ is
\begin{equation}
  L_{*}= L_R \left( \frac{M_{*}}{M_R} \right)^{1/3},
\end{equation}
where $L_{R}$ is the comoving scale of the horizon at the reheating
epoch. Thus, the mass $M_{*}$ of the PBH produced during the MD epoch
is given by
\begin{equation}
\label{mass-scale-MD}
M_{*} \simeq 6.3\times 10^{22} M_\odot \left( \frac{T_R}{\rm GeV}
\right) \left( \frac{k_{*}}{{\rm Mpc}^{-1}} \right)^{-3} .
\end{equation}
We see that small PBHs of mass, $M_{*}\sim 10^{-19}M_{\odot}$ for
example, would be produced for $T_{R}\sim 10^{6}$GeV and $k_{*}\sim
10^{16}$Mpc$^{-1}$. The condition for the PBH formation in a MD
universe is discussed in Ref.~\cite{beta-MD}, where the mass fraction
of PBHs of mass $M_{*}$ is estimated as~\cite{beta-MD}
\begin{equation}
    \beta_{*}(M_{*}) \simeq 2\times 10^{-2} \sigma(M_{*})^{13/2}.
\end{equation}
Notice that the mass fraction has a weeker dependence on $\sigma$ than
in the RD case [see Eq.(\ref{bh-frac})]

During the MD era, the mass fraction $\beta_{*}$ stays constant and
hence the density of the black holes of mass $M_{*}$,
$\rho_{BH}(M_{*})$, is given by
\begin{equation}
    \label{bh-density2} \frac{\rho_{BH}(M_{*})}{s} \simeq
    \frac{3}{4}\beta_{*} (M_{*})T_R.
\end{equation}
We can write the present density parameter $\Omega_{BH}(M_{*})$ for
the black holes of mass $M_{*}$ as
\begin{equation}
\label{Omega-MD}
   \Omega_{BH}(M_{*})h^2 \simeq 2.1\times 10^8 \beta_{*} (M_{*})
    \left(\frac{T_R}{\rm GeV} \right).
\end{equation}

\section{Double Inflation Model}

We adopt a double inflation model proposed in
Ref.\cite{dynamical-tuning}. The model consists of two inflationary
stages; the first one is called preinflation and we take a hybrid
inflation~\cite{hybrid} (see also Ref.\cite{hybrid2}) as the
preinflation. We also assume that the second inflationary stage is
realized by a new inflation model~\cite{Izawa-New-inflation} and its
$e$-fold number is smaller than $\sim 60$.  Thus, the density
fluctuations on large scales are produced during the preinflation and
their amplitude should be normalized to the COBE data~\cite{COBE}. On
the other hand, the new inflation produces fluctuations on small
scales. Since the amplitude of small scale fluctuations is free from
the COBE normalization, we expect that the new inflation can produce
large density fluctuations enough to form PBHs. We choose the
predicted power spectrum to be almost scale invariant ($n_{s} \simeq
1$) on large cosmological scales which is favored for the structure
formation of the universe~\cite{White}.  On the other hand, the new
inflation gives the power spectrum which has large amplitude and
shallow slope ($n_{s} < 1$) on small scales. Thus, this power spectrum
has a large and sharp peak on the scale corresponding to a turning
epoch from the preinflation to the new inflation, and we expect that
PBHs are produced at that scale.

As for the detailed argument of the dynamics of our model, see
Refs.\cite{KKSY-LSS,KSY-PBH,KY-PBH}.

\subsection{Preinflation}

First, let us briefly discuss a hybrid inflation model~\cite{hybrid}.
The hybrid inflation model contains two kinds of superfields: one is
$S(x,\theta)$ and the others are a pair of $\Psi(x,\theta)$ and
$\bar{\Psi}(x,\theta)$. Here $\theta$ is the Grassmann number denoting
superspace. The model is based on the U$(1)_R$ symmetry under which
$S(\theta) \rightarrow e^{2i\alpha} S(\theta e^{-i\alpha})$ and
$\Psi(\theta) \bar{\Psi}(\theta) \rightarrow \Psi(\theta e^{-i\alpha})
\bar{\Psi}(\theta e^{-i\alpha})$. The superpotential is given
by~\cite{hybrid}
\begin{equation}
\label{superpot-pre}
    W(S,\Psi,\bar{\Psi}) = -\mu^{2} S + \lambda S \bar{\Psi}\Psi.
\end{equation}
The $R$-invariant K\"ahler potential is given by
\begin{equation}
\label{kahlerpot-pre}
    K(S,\Psi,\bar{\Psi}) = |S|^{2} + |\Psi|^{2} + |\bar{\Psi}|^{2}
    + \cdots ,
\end{equation}
where the ellipsis denotes higher-order terms which we neglect in the
present analysis for simplicity. We gauge the U$(1)$ phase
rotation:$\Psi \rightarrow e^{i\delta}\Psi$ and $\bar\Psi \rightarrow
e^{-i\delta}\bar\Psi$. To satisfy the $D$-term flatness condition we
take always $\Psi = \bar\Psi$ in our analysis.

We define $N_{\rm COBE}$ as the $e$-fold number corresponding to the
COBE scale and the COBE normalization leads to a condition for the
inflaton potential,
\begin{equation}
  \label{eq:COBE-cond}
 \left| \frac{V^{3/2}}{V'} \right|_{N_{\rm COBE}}
  \simeq 5.3\times 10^{-4},
\end{equation}
where $V$ is the inflaton potential obtained from
Eqs.(\ref{superpot-pre}) and (\ref{kahlerpot-pre}). In the hybrid
inflation model, density fluctuations are almost scale invariant;
\begin{equation}
  n_{\rm pre} \simeq \left. 1+2\left( \frac{V''}{V} \right) -3\left
  ( \frac{V'}{V} \right)^2 \right|_{N_{\rm COBE}}\simeq
  1-\frac{1}{N_{\rm COBE}} \simeq 1,
\end{equation}
where $n_{\rm pre}$ is a spectral index for a power spectrum of
density fluctuations.

\subsection{New inflation}

Now, we consider a new inflation model.  We adopt an inflation model
proposed in Ref.~\cite{Izawa-New-inflation}.  The inflaton superfield
$\phi(x, \theta)$ is assumed to have an $R$ charge $2/(n+1)$ and
U$(1)_{R}$ is dynamically broken down to a discrete $Z_{2nR}$ at a
scale $v$, which generates an effective
superpotential~\cite{Izawa-New-inflation,dynamical-tuning},
\begin{equation}
        W(\phi) = v^{2}\phi - \frac{g}{n+1}\phi^{n+1}.
        \label{sup-pot2}
\end{equation}
The $R$-invariant effective K\"ahler potential is given by
\begin{equation}
    \label{new-kpot}
    K(\phi,\chi) = |\phi|^2 +\frac{\kappa}{4}|\phi|^4 
    + \cdots ,
\end{equation}
where $\kappa$ is a constant of order $1$. 
We require that supersymmetry breaking effects make the potential
energy at a vacuum vanish, and 
%
we have a relation between $v$ and the gravitino mass $m_{3/2}$ as
(for details, see Ref.~\cite{Izawa-New-inflation})
\begin{equation}
        m_{3/2} \simeq \left(\frac{n}{n+1}\right) |v|^{2}
        \left|\frac{v^{2}}{g}\right|^{\frac{1}{n}}.
        \label{gravitino-mass}
\end{equation}

The inflaton $\phi(x)$ (the scalar component of $\phi(x,\ \theta)$)
has a mass $m_{\phi}$ in the vacuum with
\begin{equation}
        m_{\phi} \simeq n |g|^{1/n}|v|^{2-2/n}.
        \label{inftaton-mass}
\end{equation}
The inflaton $\phi$ may decay into ordinary particles through
gravitationally suppressed interactions, which yields reheating
temperature $T_R$ given by\footnote{
  The decay rate of the inflaton $\phi$ is discussed in
  Ref.\cite{dynamical-tuning}
}
\begin{eqnarray}
    \label{reheat-temp}
    T_R \simeq 0.1 m_{\phi}^{3/2} &\simeq& 2.4\times 10^{17} {\rm
      GeV}n^{3/2} |g|^{3/2n}|v|^{3(1-1/n)} \\
&\lesssim& 10^{6}{\rm GeV}\quad {\rm for}\quad m_{3/2}\lesssim 1 {\rm
  TeV}, n\ge 3.
\end{eqnarray}
An important point on the above density fluctuations is that it
results in a tilted spectrum with spectral index $n_{\rm new}$ given
by (see Refs.~\cite{dynamical-tuning,Izawa-New-inflation})
\begin{equation}
    \label{eq:new-index}
    n_{\rm new} \simeq 1 - 2 \kappa.
\end{equation}

\subsection{Initial value and fluctuations of the inflaton $\varphi$}

The crucial point observed in Ref.~\cite{dynamical-tuning} is that the
preinflation sets dynamically the initial condition for the new
inflation.  We identify the inflaton field $\varphi(x)/\sqrt{2}$ with
the real part of the field $\phi(x)$. It gets an effective mass
$m_{\rm eff}\sim \mu^2$ during the
preinflation~\cite{dynamical-tuning}.  Thus, this inflaton $\varphi$
tends to the potential minimum,
\begin{equation}
    \label{deviation}
    \varphi_{\rm min} \simeq -\frac{\sqrt{2}}{\sqrt{\lambda}}
    v\left(\frac{v}{\mu}\right).
\end{equation}
Notice that $\varphi_{\rm min}$ deviates from zero due to the presence
of a linear term $v^{2}\mu^{2}S \varphi$ (see
Ref.~\cite{dynamical-tuning}).
Thus, at the end of the preinflation the $\varphi$ settles down to
this $\varphi_{\rm min}$.

After the preinflation, the universe becomes MD because of the
oscillation of the inflaton for preinflation.
During the MD era between the two inflations, the energy density
scales as $\propto a^{-3}$, and the new inflaton oscillates around
$\varphi=0$ with its amplitude decreasing as $\propto a^{-3/4}$. Since
the scale factor increases by a factor $(\mu/v)^{4/3}$ during this
era, the mean initial value $\varphi_b$ of $\varphi$ at the beginning
of the new inflation is written as
\begin{equation}
    \label{eq:init-new-inflaton}
    \varphi_b \simeq \frac{\sqrt{2}}{\sqrt{\lambda}}
    v\left(\frac{v}{\mu}\right)^2.
\end{equation}

Therefore, the amplitude of fluctuations with comoving wavelength
corresponding to the horizon scale at the beginning of the new
inflation is given by
\begin{equation}
  \delta \varphi \simeq \frac{H_{\rm pre}}{2\pi} \left(\frac{H_{\rm
        pre}}{m_{\rm eff}}\right)^{\frac{1}{2}}
  \left[ \left( \frac{\mu}{v} \right)^{2/3} \right]^{-3/2} \left [
    \left( \frac{\mu}{v} \right)^{4/3} \right]^{-3/4}
  \simeq \frac{H_{\rm pre}}{3^{1/4}2\pi}
     \left(\frac{v}{\mu}\right)^{2},  
\label{eq:q-fluctuation}
\end{equation}
where $H_{\rm pre}$ is the Hubble parameter during the hybrid inflation,
$H^{2}_{\rm pre}\simeq \mu^{4}/3$.
The fluctuations given by Eq.~(\ref{eq:q-fluctuation}) are a little
less than newly induced fluctuations at the beginning of the new
inflation [$\delta \varphi_{\rm new}\simeq v^2/(2\pi\sqrt{3}$)].
Moreover, the fluctuations produced during the preinflation are more
suppressed for smaller wavelength. Thus, we assume that the
fluctuations of $\varphi$ induced in the preinflation are negligible
compared with fluctuations produced by the new inflation.
As mentioned before, the new inflation gives the
tilted spectrum on small scales [see Eq.~(\ref{eq:new-index})] and
hence the fluctuations at the scale corresponding to the beginning of
the new inflation is dominant.

Now let us estimate $e$-fold number which corresponds to our current
horizon. The $e$-fold number is given by~\cite{Liddle-Lyth}
\begin{equation}
  N_{\rm tot} = 62 - \ln\frac{k}{a_0H_0} -\ln \frac{10^{16}{\rm
      GeV}}{V_{k}^{1/4}} +\ln\frac{V_{k}^{1/4}}{V_{\rm end}^{1/4}} -
  \frac{1}{3} \ln \frac{V_{\rm end}^{1/4}}{\rho_{\rm reh}^{1/4}},
\end{equation}
where $V_{k}$ is a potential energy when a given scale $k$ leaves the
horizon, $V_{\rm end}$ that when the inflation ends, and $\rho_{\rm
  reh}$ energy density at the time of reheating.

We take $V_{k}\simeq V_{\rm end}$, and $\rho_{\rm reh}^{1/4} \simeq
{\rm a\ few}\times T_{\rm reh}$. For $k=a_0H_0$ (i.e., the present
horizon scale), we have
\begin{equation}
N_{\rm tot} \simeq 67.1 + \left(\frac{5}{3} -\frac{1}{n} \right) \ln v 
+\frac{1}{2} \ln n +\frac{1}{2n} \ln g.
\end{equation}
In estimating $N_{\rm COBE}$ we must take into account the fact that
the fluctuations induced at $e$-fold number less than $(2/3)\ln
(\mu/v)$ before the end of the hybrid inflation reenter the horizon
before the new inflation starts. Such fluctuations are cosmologically
irrelevant since the new inflation produce much larger
fluctuations~\cite{KKSY-LSS}.  Thus, $N_{\rm COBE}$ is given by
\begin{eqnarray}
  N_{\rm COBE} &=& N_{\rm tot} -N_{\rm new} + \frac{2}{3} \ln
  \frac{\mu}{v}\nonumber\\ 
  &\simeq& 67.1 + \left(\frac{5}{3} -\frac{1}{n} \right) \ln v 
+\frac{1}{2} \ln n +\frac{1}{2n} \ln g -N_{\rm new} + \frac{2}{3} \ln
\frac{\mu}{v}.
\end{eqnarray}
The COBE normalization in Eq.~(\ref{eq:COBE-cond}) should be imposed
by using this $N_{\rm COBE}$.

\subsection{Numerical Results}

We estimate density fluctuations in the double inflation by
calculating evolution of $\varphi$ and $\sigma$ numerically. For given
parameters $\kappa$ and $\lambda$, we obtain the break scale $k_b$ and
the amplitude of density fluctuations produced at the beginning of new
inflation $\delta_{b}$. Here, $k_{b}^{-1}$ is the comoving scale
corresponding to the Hubble radius at the beginning of the new
inflation (a turning epoch). We can understand the qualitative
dependence of $(k_b, \delta_{b})$ on $(\kappa, \lambda)$ as follows:
When $\kappa$ is large, the slope of the potential for the new
inflation is too steep, and the new inflation cannot last for a long
time. Therefore, the break occurs at smaller scales. As for
$\delta_{b}$, we can see from Eq.(\ref{eq:COBE-cond}) that as
$\lambda$ gets larger, $\mu$ also gets large. In addition, we can show
that
\begin{equation}
\delta_{b}\equiv  \left( \frac{\delta \rho}{\rho} \right)_{{\rm new},
  k_{b}} \propto   \frac{\sqrt{\lambda}    \mu^2}{\kappa} \sim \frac{\lambda^{3/2}}{\kappa},
\end{equation}
for a given $v$ (see Ref.~\cite{KKSY-LSS}). Thus, we have larger
$\delta_{b}$ for larger $\lambda$.

\section{Mass Variance and Density Fluctuations}

Our double inflation model predicts the amplitude of density
fluctuations $\delta_{b}$ as a function of inflation parameters. On
the other hand, the black hole abundance is expressed as a function of
mass variance $\sigma$ at the time when the corresponding scale enters
the horizon. Therefore, when we compare the observations with the
prediction of our model, we need a relation between the mass variance
$\sigma$ and the fluctuations $\delta_{b}$.

For the power spectrum with the break scale $k_{b}^{-1}$ which enters
the horizon during the RD epoch, we have a relation between the mass
variance and the amplitude of fluctuations as\footnote{In
  Refs.~\cite{KSY-PBH,KY-PBH}, an incorrect relation $\delta_{b}\simeq
  \sigma_{b}/6$ was used.}
\begin{equation}
\label{sigma-delta-RD}
\delta_{b} \simeq \sigma_{b}/ 0.65, 
\end{equation}
(the numerical factor depends on the tilted spectral index $n_{s}$,
and within the parameter range we consider, this factor lies between
$0.62 \sim 0.67$.).
For the power spectrum with the break scale $k_{b}^{-1}$ which enters
the horizon during the MD epoch, we have a relation between the mass
variance and the amplitude of fluctuations as
\begin{equation}
\label{sigma-delta-MD}
\delta_{b} \simeq \sigma_{b}/ 2.3,
\end{equation}
(again, the numerical factor depends on the tilted spectral index
$n_{s}$, and within the parameter range we consider, this factor lies
between $2.1 \sim 2.8$.).

\begin{figure}[htbp]
  \begin{center}
    \centerline{\psfig{figure=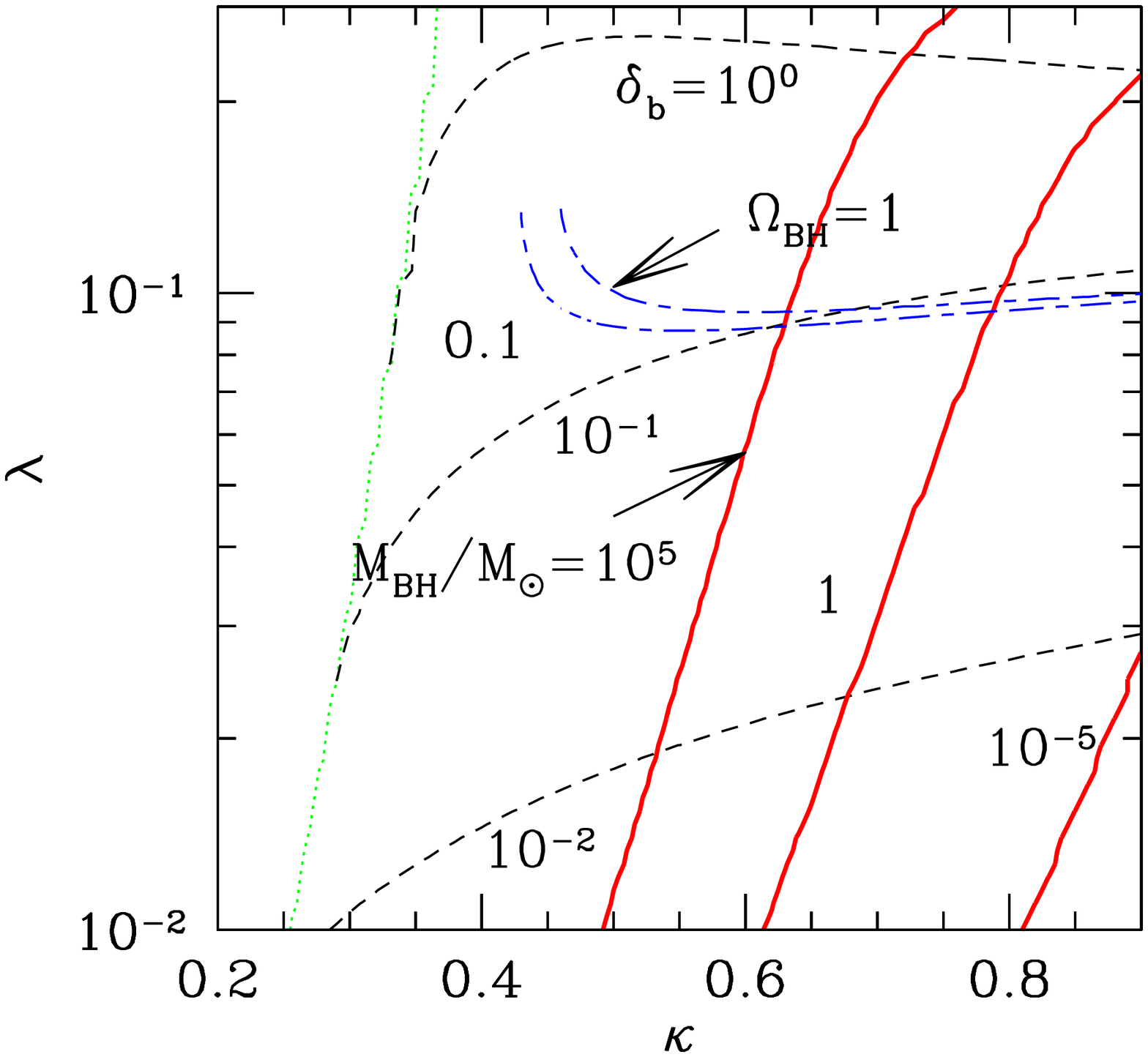,width=14cm}}
    \caption{The amplitude of density fluctuations ($\delta_{b}$) and
      PBH masses. Here we take $n=4, g=1, v=10^{-7}$. The region on
      left hand side is irrelevant since a break scale exceeds our
      current horizon. The thick solid lines correspond to $M_{\rm
        BH}/M_{\odot}=10^{5}, 1$, and $10^{-5}$, from left to right.
      The short dash - long dash lines show $\Omega_{\rm BH}\simeq 1$
      (top) and $0.1$ (bottom).  }
    \label{fig:DM-PBH}
  \end{center}
\end{figure}

First, let us consider PBH dark matter with $\Omega_{BH}\sim 1$ which
are produced during the RD epoch. Since the density fluctuations at
the break scale is dominant, and the mass fraction $\beta_{*}$ has a
sharp peak at that scale, only the PBHs of mass corresponding to the
break scale are formed.  For PBHs produced during the RD epoch (after
reheating process), we have, from Eq.~(\ref{Omega-RD-mass}),
\begin{equation}
   \Omega_{BH}h^2 \simeq 5.4\times 10^{7} \beta_{*} \left
     ( \frac{g_{*}}{50} \right)^{-1/4} \left( \frac{M_*}{M_{\odot}}
     \right)^{-1/2} .
\end{equation}
For example, if we require that the black holes with mass $\sim
M_\odot$ ($=$MACHOs) be dark matter in the present universe, i.e.
$\Omega_{BH}h^2 \sim 0.25$, we obtain $\beta_{*} \sim 5\times
10^{-9}$, and from Eq.(\ref{bh-frac}) we obtain 
\begin{equation}
    \sigma (M_{\odot}) \simeq 0.06.
\end{equation}
From Eq.(\ref{sigma-delta-RD}), we see that the break amplitude is
$\delta_{b} \simeq 0.06/0.65 \simeq 0.092$, and we find from
Eq.(\ref{mass-scale-RD}) that the break scale is $2.5\times 10^{7}{\rm
  Mpc}^{-1}$. In Fig.\ref{fig:DM-PBH}, we plot the numerical results
of our double inflation model for $n=4, g=1$, and $v=10^{-7}$. We see
a wide range of parameter space which may account for DM
($\Omega_{BH}\simeq 0.1-1$).

\begin{figure}[htbp]
  \begin{center}
    \centerline{\psfig{figure=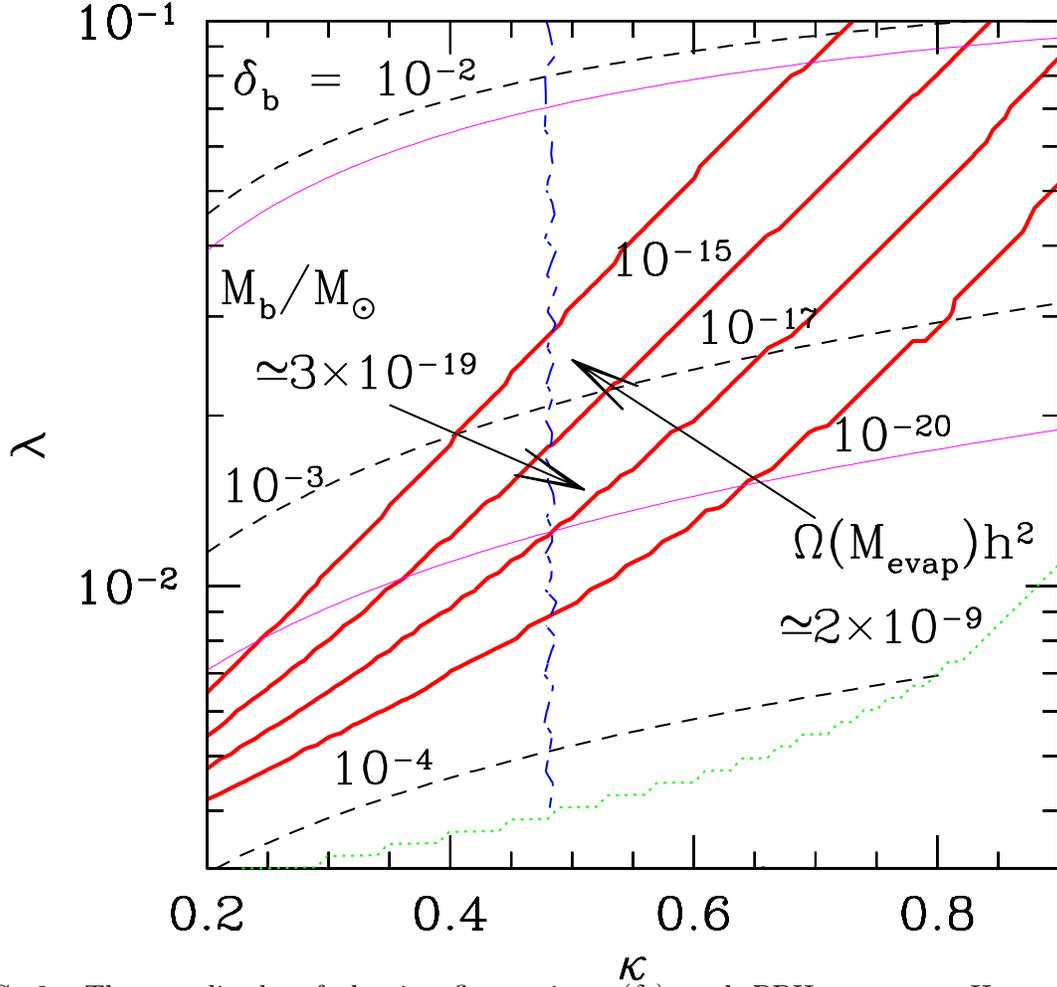,width=14cm}}
    \caption{The amplitude of density fluctuations ($\delta_{b}$) and
      PBH masses. Here we take $n=3, g=10^{-4}, v=10^{-6}$. The region
      on down right corner is irrelevant since a break scale is too
      small and does not cross horizon during inflation.  The thick
      solid lines correspond to the break masses $M_{\rm b}/M_{\odot}
      = 10^{-15}, 10^{-17}, 3\times 10^{-19}$, and $10^{-20}$, from
      left to right.  The short dash - long dash line shows $\Omega
      (M_{\rm evap})h^{2}\simeq 2\times 10^{-9}$. The two thin solid
      lines show the region $\Omega (M_{b})h^{2}\simeq 1$ (top) and
      $\simeq 2\times 10^{-9}$ (bottom), respectively.}
    \label{fig:BESS-PBH}
  \end{center}
\end{figure}

Next, let us consider another interesting mass range of PBHs which are
evaporating now ($M_{\rm evap}\sim 3 \times 10^{-19} M_\odot$). The
PBHs of such light mass are produced during the MD epoch and from
Eq.(\ref{Omega-MD}) we find
\begin{equation}
    \Omega_{BH}(M_{*})h^2 \simeq 4.2\times 10^6 \sigma(M_{*})^{13/2} \left(
\frac{T_R}{\rm GeV} \right).
\end{equation}
It has been reported, recently, that the BESS experiment~\cite{BESS}
has observed antiproton fluxes, which may be explained by the
evaporation of PBHs if $\Omega_{BH}h^2 \simeq 2\times
10^{-9}$~\cite{BESS}. In order to explain the BESS result by
evaporating PBHs we need
\begin{equation}
\sigma(M_{\rm evap}) \simeq 4.4\times 10^{-3} \left( \frac{T_{R}}{\rm 
    GeV} \right)^{-2/13}.
\end{equation}
From Eq.(\ref{reheat-temp}), we estimate the required fraction of the
evaporating PBHs as
\begin{equation}
\sigma(M_{\rm evap}) \simeq 9.3\times 10^{-6} n^{-3/13} |g|^{-3/13n}
v^{-6(1-1/n)/13}.
\end{equation}
Since the mass variance $\sigma(M)$ scales as $\sigma(M)\propto
M^{(1-n_s)/6}$~\cite{Green} during the MD epoch, we obtain the mass
variance at the break scale as
\begin{equation}
\sigma_b \simeq 9.3\times 10^{-6} n^{-3/13} |g|^{-3/13n}
v^{-6(1-1/n)/13}\left( \frac{M_{b}}{M_{\rm evap}} \right)^{(1-n_s)/6} ,
\end{equation}
and the amplitude $\delta_b$ is $\delta_b \simeq \sigma_b/2.3$ and
$n_s \simeq 1-2\kappa$. In Fig.\ref{fig:BESS-PBH}, we plot an example
of the numerical results of our double inflation model\footnote{For
  the case of $n=4$, we do not find a consistent parameter region with
  the BESS experiment.} for $n=3, g=10^{-4}$, and $v=10^{-6}$. As
shown in the figure, we have a set of inflation parameters $(\kappa,
\lambda$) which may account for the BESS experiment.

\section{Conclusions and Discussions}

In this paper we have studied the formation of PBHs by taking a double
inflation model in supergravity.  We have shown that in a wide range
of parameter space PBHs are produced of various masses. These PBHs are
interesting since, for example, they may be identified with MACHOs
($M\sim M_{\odot}$) in the halo of our galaxy.  Or, they may be PBHs
which are just evaporating now ($M\sim 10^{-19}M_{\odot}$). Such black
holes are one of the interesting candidates for the sources of
antiproton fluxes recently observed in the BESS detector~\cite{BESS}.

The dark matter PBHs play a role of the cold dark matter on the large
scale structure formation.  The scales of the fluctuations for PBH
formation themselves are much smaller than the galactic scale and thus
we cannot see any signals for such fluctuations in $\delta T/T$
measurements.  However, the PBHs may be a source of gravitational
waves.  If the PBHs dominate dark matter of the present universe, some
of them likely form binaries.  Such binary black holes coalesce and
produce significant gravitational waves~\cite{Nakamura} which may be
observable in future detectors.

\section*{Acknowledgment}

T. K. is grateful to K. Sato for his continuous encouragement. A part
of work is supported by Grant-in-Aid of the Ministry of Education and
by Grant-in-Aid, Priority Area ``Supersymmetry and Unified Theory of
Elementary Particles'' (\#707).


\end{document}